\documentclass[a4paper,11pt]{article}
\usepackage{color}
\usepackage[dvipsnames, table]{xcolor}
\usepackage{pos}
\usepackage{enumerate}
\usepackage{tabularx}

\usepackage[breakable]{tcolorbox}
\usepackage{fancyvrb} 

\usepackage{textcomp} 
\AtBeginDocument{%
    \def\PYZsq{\textquotesingle}
}
\usepackage{upquote} 

\makeatletter
\def\PY@reset{\let\PY@it=\relax \let\PY@bf=\relax%
    \let\PY@ul=\relax \let\PY@tc=\relax%
    \let\PY@bc=\relax \let\PY@ff=\relax}
\def\PY@tok#1{\csname PY@tok@#1\endcsname}
\def\PY@toks#1+{\ifx\relax#1\empty\else%
    \PY@tok{#1}\expandafter\PY@toks\fi}
\def\PY@do#1{\PY@bc{\PY@tc{\PY@ul{%
    \PY@it{\PY@bf{\PY@ff{#1}}}}}}}
\def\PY#1#2{\PY@reset\PY@toks#1+\relax+\PY@do{#2}}

\expandafter\def\csname PY@tok@w\endcsname{\def\PY@tc##1{\textcolor[rgb]{0.73,0.73,0.73}{##1}}}
\expandafter\def\csname PY@tok@c\endcsname{\let\PY@it=\textit\def\PY@tc##1{\textcolor[rgb]{0.25,0.50,0.50}{##1}}}
\expandafter\def\csname PY@tok@cp\endcsname{\def\PY@tc##1{\textcolor[rgb]{0.74,0.48,0.00}{##1}}}
\expandafter\def\csname PY@tok@k\endcsname{\let\PY@bf=\textbf\def\PY@tc##1{\textcolor[rgb]{0.00,0.50,0.00}{##1}}}
\expandafter\def\csname PY@tok@kp\endcsname{\def\PY@tc##1{\textcolor[rgb]{0.00,0.50,0.00}{##1}}}
\expandafter\def\csname PY@tok@kt\endcsname{\def\PY@tc##1{\textcolor[rgb]{0.69,0.00,0.25}{##1}}}
\expandafter\def\csname PY@tok@o\endcsname{\def\PY@tc##1{\textcolor[rgb]{0.40,0.40,0.40}{##1}}}
\expandafter\def\csname PY@tok@ow\endcsname{\let\PY@bf=\textbf\def\PY@tc##1{\textcolor[rgb]{0.67,0.13,1.00}{##1}}}
\expandafter\def\csname PY@tok@nb\endcsname{\def\PY@tc##1{\textcolor[rgb]{0.00,0.50,0.00}{##1}}}
\expandafter\def\csname PY@tok@nf\endcsname{\def\PY@tc##1{\textcolor[rgb]{0.00,0.00,1.00}{##1}}}
\expandafter\def\csname PY@tok@nc\endcsname{\let\PY@bf=\textbf\def\PY@tc##1{\textcolor[rgb]{0.00,0.00,1.00}{##1}}}
\expandafter\def\csname PY@tok@nn\endcsname{\let\PY@bf=\textbf\def\PY@tc##1{\textcolor[rgb]{0.00,0.00,1.00}{##1}}}
\expandafter\def\csname PY@tok@ne\endcsname{\let\PY@bf=\textbf\def\PY@tc##1{\textcolor[rgb]{0.82,0.25,0.23}{##1}}}
\expandafter\def\csname PY@tok@nv\endcsname{\def\PY@tc##1{\textcolor[rgb]{0.10,0.09,0.49}{##1}}}
\expandafter\def\csname PY@tok@no\endcsname{\def\PY@tc##1{\textcolor[rgb]{0.53,0.00,0.00}{##1}}}
\expandafter\def\csname PY@tok@nl\endcsname{\def\PY@tc##1{\textcolor[rgb]{0.63,0.63,0.00}{##1}}}
\expandafter\def\csname PY@tok@ni\endcsname{\let\PY@bf=\textbf\def\PY@tc##1{\textcolor[rgb]{0.60,0.60,0.60}{##1}}}
\expandafter\def\csname PY@tok@na\endcsname{\def\PY@tc##1{\textcolor[rgb]{0.49,0.56,0.16}{##1}}}
\expandafter\def\csname PY@tok@nt\endcsname{\let\PY@bf=\textbf\def\PY@tc##1{\textcolor[rgb]{0.00,0.50,0.00}{##1}}}
\expandafter\def\csname PY@tok@nd\endcsname{\def\PY@tc##1{\textcolor[rgb]{0.67,0.13,1.00}{##1}}}
\expandafter\def\csname PY@tok@s\endcsname{\def\PY@tc##1{\textcolor[rgb]{0.73,0.13,0.13}{##1}}}
\expandafter\def\csname PY@tok@sd\endcsname{\let\PY@it=\textit\def\PY@tc##1{\textcolor[rgb]{0.73,0.13,0.13}{##1}}}
\expandafter\def\csname PY@tok@si\endcsname{\let\PY@bf=\textbf\def\PY@tc##1{\textcolor[rgb]{0.73,0.40,0.53}{##1}}}
\expandafter\def\csname PY@tok@se\endcsname{\let\PY@bf=\textbf\def\PY@tc##1{\textcolor[rgb]{0.73,0.40,0.13}{##1}}}
\expandafter\def\csname PY@tok@sr\endcsname{\def\PY@tc##1{\textcolor[rgb]{0.73,0.40,0.53}{##1}}}
\expandafter\def\csname PY@tok@ss\endcsname{\def\PY@tc##1{\textcolor[rgb]{0.10,0.09,0.49}{##1}}}
\expandafter\def\csname PY@tok@sx\endcsname{\def\PY@tc##1{\textcolor[rgb]{0.00,0.50,0.00}{##1}}}
\expandafter\def\csname PY@tok@m\endcsname{\def\PY@tc##1{\textcolor[rgb]{0.40,0.40,0.40}{##1}}}
\expandafter\def\csname PY@tok@gh\endcsname{\let\PY@bf=\textbf\def\PY@tc##1{\textcolor[rgb]{0.00,0.00,0.50}{##1}}}
\expandafter\def\csname PY@tok@gu\endcsname{\let\PY@bf=\textbf\def\PY@tc##1{\textcolor[rgb]{0.50,0.00,0.50}{##1}}}
\expandafter\def\csname PY@tok@gd\endcsname{\def\PY@tc##1{\textcolor[rgb]{0.63,0.00,0.00}{##1}}}
\expandafter\def\csname PY@tok@gi\endcsname{\def\PY@tc##1{\textcolor[rgb]{0.00,0.63,0.00}{##1}}}
\expandafter\def\csname PY@tok@gr\endcsname{\def\PY@tc##1{\textcolor[rgb]{1.00,0.00,0.00}{##1}}}
\expandafter\def\csname PY@tok@ge\endcsname{\let\PY@it=\textit}
\expandafter\def\csname PY@tok@gs\endcsname{\let\PY@bf=\textbf}
\expandafter\def\csname PY@tok@gp\endcsname{\let\PY@bf=\textbf\def\PY@tc##1{\textcolor[rgb]{0.00,0.00,0.50}{##1}}}
\expandafter\def\csname PY@tok@go\endcsname{\def\PY@tc##1{\textcolor[rgb]{0.53,0.53,0.53}{##1}}}
\expandafter\def\csname PY@tok@gt\endcsname{\def\PY@tc##1{\textcolor[rgb]{0.00,0.27,0.87}{##1}}}
\expandafter\def\csname PY@tok@err\endcsname{\def\PY@bc##1{\setlength{\fboxsep}{0pt}\fcolorbox[rgb]{1.00,0.00,0.00}{1,1,1}{\strut ##1}}}
\expandafter\def\csname PY@tok@kc\endcsname{\let\PY@bf=\textbf\def\PY@tc##1{\textcolor[rgb]{0.00,0.50,0.00}{##1}}}
\expandafter\def\csname PY@tok@kd\endcsname{\let\PY@bf=\textbf\def\PY@tc##1{\textcolor[rgb]{0.00,0.50,0.00}{##1}}}
\expandafter\def\csname PY@tok@kn\endcsname{\let\PY@bf=\textbf\def\PY@tc##1{\textcolor[rgb]{0.00,0.50,0.00}{##1}}}
\expandafter\def\csname PY@tok@kr\endcsname{\let\PY@bf=\textbf\def\PY@tc##1{\textcolor[rgb]{0.00,0.50,0.00}{##1}}}
\expandafter\def\csname PY@tok@bp\endcsname{\def\PY@tc##1{\textcolor[rgb]{0.00,0.50,0.00}{##1}}}
\expandafter\def\csname PY@tok@fm\endcsname{\def\PY@tc##1{\textcolor[rgb]{0.00,0.00,1.00}{##1}}}
\expandafter\def\csname PY@tok@vc\endcsname{\def\PY@tc##1{\textcolor[rgb]{0.10,0.09,0.49}{##1}}}
\expandafter\def\csname PY@tok@vg\endcsname{\def\PY@tc##1{\textcolor[rgb]{0.10,0.09,0.49}{##1}}}
\expandafter\def\csname PY@tok@vi\endcsname{\def\PY@tc##1{\textcolor[rgb]{0.10,0.09,0.49}{##1}}}
\expandafter\def\csname PY@tok@vm\endcsname{\def\PY@tc##1{\textcolor[rgb]{0.10,0.09,0.49}{##1}}}
\expandafter\def\csname PY@tok@sa\endcsname{\def\PY@tc##1{\textcolor[rgb]{0.73,0.13,0.13}{##1}}}
\expandafter\def\csname PY@tok@sb\endcsname{\def\PY@tc##1{\textcolor[rgb]{0.73,0.13,0.13}{##1}}}
\expandafter\def\csname PY@tok@sc\endcsname{\def\PY@tc##1{\textcolor[rgb]{0.73,0.13,0.13}{##1}}}
\expandafter\def\csname PY@tok@dl\endcsname{\def\PY@tc##1{\textcolor[rgb]{0.73,0.13,0.13}{##1}}}
\expandafter\def\csname PY@tok@s2\endcsname{\def\PY@tc##1{\textcolor[rgb]{0.73,0.13,0.13}{##1}}}
\expandafter\def\csname PY@tok@sh\endcsname{\def\PY@tc##1{\textcolor[rgb]{0.73,0.13,0.13}{##1}}}
\expandafter\def\csname PY@tok@s1\endcsname{\def\PY@tc##1{\textcolor[rgb]{0.73,0.13,0.13}{##1}}}
\expandafter\def\csname PY@tok@mb\endcsname{\def\PY@tc##1{\textcolor[rgb]{0.40,0.40,0.40}{##1}}}
\expandafter\def\csname PY@tok@mf\endcsname{\def\PY@tc##1{\textcolor[rgb]{0.40,0.40,0.40}{##1}}}
\expandafter\def\csname PY@tok@mh\endcsname{\def\PY@tc##1{\textcolor[rgb]{0.40,0.40,0.40}{##1}}}
\expandafter\def\csname PY@tok@mi\endcsname{\def\PY@tc##1{\textcolor[rgb]{0.40,0.40,0.40}{##1}}}
\expandafter\def\csname PY@tok@il\endcsname{\def\PY@tc##1{\textcolor[rgb]{0.40,0.40,0.40}{##1}}}
\expandafter\def\csname PY@tok@mo\endcsname{\def\PY@tc##1{\textcolor[rgb]{0.40,0.40,0.40}{##1}}}
\expandafter\def\csname PY@tok@ch\endcsname{\let\PY@it=\textit\def\PY@tc##1{\textcolor[rgb]{0.25,0.50,0.50}{##1}}}
\expandafter\def\csname PY@tok@cm\endcsname{\let\PY@it=\textit\def\PY@tc##1{\textcolor[rgb]{0.25,0.50,0.50}{##1}}}
\expandafter\def\csname PY@tok@cpf\endcsname{\let\PY@it=\textit\def\PY@tc##1{\textcolor[rgb]{0.25,0.50,0.50}{##1}}}
\expandafter\def\csname PY@tok@c1\endcsname{\let\PY@it=\textit\def\PY@tc##1{\textcolor[rgb]{0.25,0.50,0.50}{##1}}}
\expandafter\def\csname PY@tok@cs\endcsname{\let\PY@it=\textit\def\PY@tc##1{\textcolor[rgb]{0.25,0.50,0.50}{##1}}}

\def\PYZbs{\char`\\}
\def\PYZus{\char`\_}
\def\PYZob{\char`\{}
\def\PYZcb{\char`\}}

\def\PYZgt{\char`\>}

\def\PYZdl{\char`\$}
\def\PYZhy{\char`\-}
\def\PYZsq{\char`\'}
\def\PYZdq{\char`\"}

\makeatother

    \makeatletter
        \newbox\Wrappedcontinuationbox
        \newbox\Wrappedvisiblespacebox
        \newcommand*\Wrappedvisiblespace {\textcolor{red}{\textvisiblespace}}
        \newcommand*\Wrappedcontinuationsymbol {\textcolor{red}{\llap{\tiny$\m@th\hookrightarrow$}}}
        \newcommand*\Wrappedcontinuationindent {3ex }
        \newcommand*\Wrappedafterbreak {\kern\Wrappedcontinuationindent\copy\Wrappedcontinuationbox}
        \newcommand*\Wrappedbreaksatspecials {%
            \def\PYGZus{\discretionary{\char`\_}{\Wrappedafterbreak}{\char`\_}}%
            \def\PYGZob{\discretionary{}{\Wrappedafterbreak\char`\{}{\char`\{}}%
            \def\PYGZcb{\discretionary{\char`\}}{\Wrappedafterbreak}{\char`\}}}%
            \def\PYGZca{\discretionary{\char`\^}{\Wrappedafterbreak}{\char`\^}}%
            \def\PYGZam{\discretionary{\char`\&}{\Wrappedafterbreak}{\char`\&}}%
            \def\PYGZlt{\discretionary{}{\Wrappedafterbreak\char`\<}{\char`\<}}%
            \def\PYGZgt{\discretionary{\char`\>}{\Wrappedafterbreak}{\char`\>}}%
            \def\PYGZsh{\discretionary{}{\Wrappedafterbreak\char`\#}{\char`\#}}%
            \def\PYGZpc{\discretionary{}{\Wrappedafterbreak\char`\%}{\char`\%}}%
            \def\PYGZdl{\discretionary{}{\Wrappedafterbreak\char`\$}{\char`\$}}%
            \def\PYGZhy{\discretionary{\char`\-}{\Wrappedafterbreak}{\char`\-}}%
            \def\PYGZsq{\discretionary{}{\Wrappedafterbreak\textquotesingle}{\textquotesingle}}%
            \def\PYGZdq{\discretionary{}{\Wrappedafterbreak\char`\"}{\char`\"}}%
            \def\PYGZti{\discretionary{\char`\~}{\Wrappedafterbreak}{\char`\~}}%
        }
        \newcommand*\Wrappedbreaksatpunct {%
            \lccode`\~`\.\lowercase{\def~}{\discretionary{\hbox{\char`\.}}{\Wrappedafterbreak}{\hbox{\char`\.}}}%
            \lccode`\~`\,\lowercase{\def~}{\discretionary{\hbox{\char`\,}}{\Wrappedafterbreak}{\hbox{\char`\,}}}%
            \lccode`\~`\;\lowercase{\def~}{\discretionary{\hbox{\char`\;}}{\Wrappedafterbreak}{\hbox{\char`\;}}}%
            \lccode`\~`\:\lowercase{\def~}{\discretionary{\hbox{\char`\:}}{\Wrappedafterbreak}{\hbox{\char`\:}}}%
            \lccode`\~`\?\lowercase{\def~}{\discretionary{\hbox{\char`\?}}{\Wrappedafterbreak}{\hbox{\char`\?}}}%
            \lccode`\~`\!\lowercase{\def~}{\discretionary{\hbox{\char`\!}}{\Wrappedafterbreak}{\hbox{\char`\!}}}%
            \lccode`\~`\/\lowercase{\def~}{\discretionary{\hbox{\char`\/}}{\Wrappedafterbreak}{\hbox{\char`\/}}}%
            \catcode`\.\active
            \catcode`\,\active
            \catcode`\;\active
            \catcode`\:\active
            \catcode`\?\active
            \catcode`\!\active
            \catcode`\/\active
            \lccode`\~`\~
        }
    \makeatother

    \let\OriginalVerbatim=\Verbatim
    \makeatletter
    \renewcommand{\Verbatim}[1][1]{%
        \sbox\Wrappedcontinuationbox {\Wrappedcontinuationsymbol}%
        \sbox\Wrappedvisiblespacebox {\FV@SetupFont\Wrappedvisiblespace}%
        \def\FancyVerbFormatLine ##1{\hsize\linewidth
            \vtop{\raggedright\hyphenpenalty\z@\exhyphenpenalty\z@
                \doublehyphendemerits\z@\finalhyphendemerits\z@
                \strut ##1\strut}%
        }%
        \def\FV@Space {%
            \nobreak\hskip\z@ plus\fontdimen3\font minus\fontdimen4\font
            \discretionary{\copy\Wrappedvisiblespacebox}{\Wrappedafterbreak}
            {\kern\fontdimen2\font}%
        }%

        \Wrappedbreaksatspecials
        \OriginalVerbatim[#1,codes*=\Wrappedbreaksatpunct]%
    }
    \makeatother

    \definecolor{incolor}{HTML}{303F9F}
    \definecolor{outcolor}{HTML}{D84315}
    \definecolor{cellborder}{HTML}{CFCFCF}
    \definecolor{cellbackground}{HTML}{F7F7F7}

    \makeatletter
    \newcommand{\boxspacing}{\kern\kvtcb@left@rule\kern\kvtcb@boxsep}
    \makeatother
    \newcommand{\prompt}[4]{
        {\ttfamily\llap{{\color{#2}#3:\hspace{3pt}#4}}\vspace{-\baselineskip}}
    }

\providecommand{\tightlist}{%
    \setlength{\itemsep}{0pt}\setlength{\parskip}{0pt}}

\title{\texttt{smelli} -- the \underline{SME}FT \underline{L}ike\underline{li}hood}

\author*[a]{Peter Stangl}

\affiliation[a]{Albert Einstein Center for Fundamental Physics,
Institute for Theoretical Physics,
University of Bern,\\
Sidlerstrasse 5, CH-3012 Bern,
Switzerland}

\emailAdd{stangl@itp.unibe.ch}

\abstract{I present the  Python package \texttt{smelli} that implements a global likelihood function in the space of dimension-six Wilson coefficients in the Standard Model Effective Field Theory (SMEFT).
The likelihood includes contributions from a large number of flavor and other precision observables, currently 399 in total.
}

\FullConference{%
  Tools for High Energy Physics and Cosmology - TOOLS2020\\
  2-6 November, 2020\\
  Institut de Physique des 2 Infinis (IP2I), Lyon, France}

 \tableofcontents

\begin{document}
\maketitle

\section{Introduction}
The Standard Model (SM) of particle physics is an extremely successful model.
However, there are several experimental as well as theoretical indications for new physics (NP) beyond the SM.
Whether a given NP scenario describes the experimental data better than the SM can be conveniently quantified by the ratio of the NP likelihood~$L_{\rm NP}$ and the SM likelihood~$L_{\rm SM}$ or, equivalently, by the difference of the log-likelihoods
\begin{equation}\label{eq:delta_log_likelihood}
 \Delta \log{L} = \log{L_{\rm NP}} - \log{L_{\rm SM}}\,.
\end{equation}
These likelihood functions are constructed from a set of measured observables and take into account uncertainties and correlations from both the measurements and the theoretical predictions.

A set of observables for which certain NP scenarios can describe the experimental data considerably better than the SM have been found e.g.\ in $B$ meson decays.
These so-called $B$~anomalies correspond to deviations from the SM predictions in measurements of neutral current $b\to s\ell\ell$ and charged current $b\to c\ell\nu$ transitions.
In particular, deviations have been found in
\begin{enumerate}[{\em (i)}]
 \item\label{anom:ang_obs} angular observables of $B\to K^*\mu^+\mu^-$~\cite{Aaij:2020nrf,Khachatryan:2015isa,ATLAS:2017dlm,Sirunyan:2017dhj,Bp_Kpstarmumu_LHCb_preliminary},

 \item\label{anom:br} branching ratios of $B\to K\mu^+\mu^-$, $B\to K^*\mu^+\mu^-$, and $B_s\to \phi\mu^+\mu^-$~\cite{Aaij:2014pli,Aaij:2015esa,Aaij:2016flj},

 \item\label{anom:rk_rkstar}
 the lepton flavor universality (LFU) observables $R_{K^{(*)}}$~\cite{Aaij:2017vbb,Aaij:2019wad,Abdesselam:2019wac,Abdesselam:2019lab}, which are $\mu/e$ ratios of ${B\to K^{(*)}\ell^+\ell^-}$ branching ratios,
 %

 \item\label{anom:Bsmumu} the branching ratio of $B_s\to \mu^+\mu^-$~\cite{Chatrchyan:2013bka,CMS:2014xfa,Aaij:2017vad,Aaboud:2018mst,LHCb:2020zud},

 \item\label{anom:rd_rdstar}
 the LFU observables $R_{D^{(*)}}$~\cite{Lees:2012xj,Lees:2013uzd,Huschle:2015rga,Sato:2016svk,Hirose:2016wfn,Aaij:2015yra,Aaij:2017uff,Abdesselam:2019dgh}, which are $\tau/e$ and $\tau/\mu$ ratios of $B\to D^{(*)}\ell\nu$ branching ratios.
 %
\end{enumerate}
While \emph{(\ref{anom:ang_obs})} and \emph{(\ref{anom:br})} could be afflicted by underestimated hadronic uncertainties, the observables in \emph{(\ref{anom:rk_rkstar})}, \emph{(\ref{anom:Bsmumu})}, and \emph{(\ref{anom:rd_rdstar})} are theoretically clean probes of NP~\cite{Bordone:2016gaq,Beneke:2017vpq,Bordone:2019vic}.
Considering the above $B$-decay observables and parameterizing NP in $b\to s\ell\ell$ and $b\to c\ell\nu$ transitions in terms of Wilson coefficients in the Weak Effective Theory~(WET), simple one- and two-parameter scenarios show a sizable $\Delta \log{L}\sim 20$ (cf.\ e.g.~\cite{Aebischer:2019mlg,Alguero:2019ptt,Datta:2019zca,Kowalska:2019ley,Hurth:2020rzx,Ciuchini:2020gvn}).

These intriguing hints for NP have led to extensive model building.
In the process, important insights have been gained:
\begin{itemize}
 \item The fact that NP above the electroweak (EW) scale has to respect SM gauge invariance leads to important correlations between low-energy observables. For example, explanations of $R_{D^{(*)}}$ in terms of left-handed contributions to $b\to c\tau\nu$  imply also contributions to $b\to s\nu\nu$, which are constrained by  $B\to K^{(*)} \nu\bar\nu$~\cite{Buras:2014fpa}.
 \item One-loop contributions can have very important effects.
 This has been observed in models explaining $R_{D^{(*)}}$ and $R_{K^{(*)}}$ using mostly 3rd generation couplings. They actually modify $\tau$ and $Z$ decays at one loop, which leads to strong constraints~\cite{Feruglio:2017rjo}. Another example is provided by models explaining $R_{D^{(*)}}$ using a contribution to semi-tauonic operators, which generate an effect in $b\to s\ell\ell$ at one loop~\cite{Bobeth:2011st,Crivellin:2018yvo}.
\end{itemize}
Essentially every model that explains some of the $B$ anomalies predicts deviations from the SM also in other observables.
In many cases, this leads to strong constraints or exclusion of a model.
So phenomenological analyses that consider only a small set of observables or neglect one-loop contributions are in many cases not sufficient to show that a given model agrees with experimental data better than the SM.
In order to show this, it is in general necessary to
\begin{itemize}
 \item compute \emph{all relevant observables} $\vec O(\vec\xi)$ (flavor observables, EW precision observables (EWPO), etc.) in terms of the Lagrangian parameters $\vec \xi$ of a NP model,
 \item take into account loop effects when computing the observables,
 \item compare the theory predictions to experimental data by constructing the NP likelihood $L_{\rm NP}$.
\end{itemize}
Performing these steps again and again for each single model one wants to analyze is a tedious task.
Fortunately, analyses of NP models can be tremendously simplified by making use of the SM effective field theory (SMEFT) in an intermediate step.

\section{The SMEFT Likelihood}
Assuming that the scale of NP $\Lambda_{\rm NP}$ is considerably larger than the EW scale and EW symmetry breaking is realized linearly, the NP effects in a given observable can be expressed in terms of the Wilson coefficients $C_i$ of the SMEFT, which are defined by the SMEFT Lagrangian~\cite{Buchmuller:1985jz,Grzadkowski:2010es}
\begin{equation}
 \mathcal{L}_{\rm SMEFT} = \mathcal{L}_{\rm SM} +\sum_{n>4}\sum_i\frac{C_i}{\Lambda_{\rm NP}^{n-4}}\,\mathcal{O}_i\,,
\end{equation}
where $\mathcal{O}_i$ are local SM gauge invariant operators constructed from the SM fields and $n$ is their canonical dimension.

The SMEFT is a powerful tool since it can connect the model building at the high scale $\Lambda_{\rm NP}$ to the phenomenology at lower scales without the need to compute hundreds of observables in each model.
A phenomenological analysis can be split into
\begin{itemize}
 \item a model-dependent part that consists of matching the NP model to the SMEFT at the scale~$\Lambda_{\rm NP}$,
 \item the model-independent phenomenology, which corresponds to
 \begin{itemize}
  \item running down the Wilson coefficients $\vec C$ from $\Lambda_{\rm NP}$ to the low scale at which the observables are computed,
  \item predicting all the relevant observables $\vec O(\vec C)$ in terms of the Wilson coefficients $\vec C$,
  \item constructing the NP likelihood $L_{\rm NP}(\vec O(\vec C))$ that compares the predictions to experimental measurements,
  \item computing $\Delta \log{L}$ using eq.~\eqref{eq:delta_log_likelihood} in order to compare the NP model to the SM.
 \end{itemize}
\end{itemize}
While it might be preferable to perform the model-dependent matching at one-loop, a large number of important one-loop effects is actually already included by the model-independent renormalization group~(RG) running and mixing in the SMEFT.


Using the above procedure, a SMEFT likelihood function $L_{\rm NP}(\vec C)$ can tremendously simplify analyses of NP models.
Many likelihood functions in the SMEFT have been considered in the literature (see e.g.~\cite{Efrati:2015eaa,Falkowski:2017pss,Alioli:2017ces,Gonzalez-Alonso:2018omy,Falkowski:2015jaa,Bobeth:2015zqa,Falkowski:2015krw,Biekotter:2018rhp,Hartland:2019bjb,Brivio:2019ius,Bissmann:2019gfc,vanBeek:2019evb,Durieux:2019rbz,Falkowski:2020pma,Ellis:2020unq}).
However, most of them are constructed from observables in one or few specific sectors, like EWPO, Higgs physics, top physics, $B$ physics, or lepton flavor violating observables.
But as discussed above, NP models generically predict new effects in several observables of various sectors.
Furthermore, SMEFT operators belonging to different sectors mix under renormalization.
Consequently, to test a NP model, the sectors should not be considered separately.
It is in fact necessary to construct the \emph{global} SMEFT likelihood, taking into account as many observables from as many sectors as possible.

\section{The \texttt{smelli} Python package}
In~\cite{Aebischer:2018iyb}, we have started constructing a global SMEFT likelihood that is provided by the Python package \texttt{smelli} (\underline{SME}FT \underline{l}ike\underline{li}hood).
It is based on
\begin{itemize}
 \item the Python package \texttt{flavio}~\cite{Straub:2018kue} that can compute hundreds of flavor and other precision observables in and beyond the SM, while properly accounting for theory uncertainties,
 \item the Wilson coefficient exchange format (WCxf)~\cite{Aebischer:2017ugx} that is used to represent and exchange large sets of Wilson coefficients in various EFTs and bases,
 \item the Python package \texttt{wilson}~\cite{Aebischer:2018bkb} that performs the RG evolution in the SMEFT and the WET as well as the matching between them.
\end{itemize}
\texttt{smelli} is built upon these tools and implements a SMEFT likelihood function constructed from currently 399 observables.
In particular, it includes
\begin{itemize}
  \tightlist
 \item flavor-changing neutral current $B$ decays,
 \item lepton flavor universality tests in charged- and neutral-current $B$ and $K$ decays,
 \item meson-antimeson mixing in the $K$, $B$, and $D$ systems,
 \item charged lepton flavor violating $B$, tau, and muon decays,
 \item the anomalous magnetic moments of the electron, muon, and tau,
 \item $Z$ and $W$ pole EWPO,
 \item nuclear and neutron beta decays,
 \item Higgs signal strengths.
\end{itemize}
%
%
Given any combination of SMEFT or WET Wilson coefficients, \texttt{smelli} computes the $\Delta \log{L}$ for each of the above sectors and then sums all of them to obtain the global $\Delta \log{L}$.
%

The \emph{full global} likelihood is work in progress and the development is open to everyone. The open-source code of \texttt{smelli} is available at \url{https://github.com/smelli/smelli}.

\subsection{Installation}
\noindent
The requirements for \texttt{smelli} are a working installation of Python version 3.5 or above and the Python package manger \texttt{pip}.
If both are present, \texttt{smelli} can be installed from the command line by entering
\begin{enumerate}[\indent {}]
    \item \textbf{\texttt{python3\ -m\ pip\ install\ smelli\ -\/-user}}
\end{enumerate}
This will download \texttt{smelli} and all its dependencies from the Python package archive (PyPI) and install it in the user's home directory without requiring root privileges (due to the option \texttt{-\/-user}).

\subsection{Using \texttt{smelli}}
\noindent
Like any Python package, \texttt{smelli} can be used
\begin{itemize}
\tightlist
 \item as a library imported from other scripts,
 \item directly in the command line interpreter,
 \item in an interactive session, e.g.\ in a Jupyter notebook.
\end{itemize}
%
How to use \texttt{smelli} is demonstrated in the following with examples from an interactive Jupyter notebook.
This notebook is available at \url{https://github.com/peterstangl/smelli-talk}.
For further information on the features of \texttt{smelli}, see~\cite{Aebischer:2018iyb} and the API documentation at \url{https://smelli.github.io}.

\subsubsection{Instantiating the likelihood}
\noindent
The main functionality of \texttt{smelli} is provided by the \texttt{GlobalLikelihood} class. It is imported by

\begin{tcolorbox}[breakable, size=fbox, boxrule=1pt, pad at break*=1mm,colback=cellbackground, colframe=cellborder]
\prompt{In}{incolor}{In}{\boxspacing}
\begin{Verbatim}[commandchars=\\\{\}]
\PY{k+kn}{from} \PY{n+nn}{smelli} \PY{k+kn}{import} \PY{n}{GlobalLikelihood}
\end{Verbatim}
\end{tcolorbox}

\noindent
If the \texttt{GlobalLikelihood} class is instantiated without any argument,

\begin{tcolorbox}[breakable, size=fbox, boxrule=1pt, pad at break*=1mm,colback=cellbackground, colframe=cellborder]
\prompt{In}{incolor}{In}{\boxspacing}
\begin{Verbatim}[commandchars=\\\{\}]
\PY{n}{gl} \PY{o}{=} \PY{n}{GlobalLikelihood}\PY{p}{(}\PY{p}{)}
\end{Verbatim}
\end{tcolorbox}

\pagebreak

\noindent
the likelihood is defined in the space of SMEFT Wilson coefficients in the \texttt{Warsaw} basis (for details on the specifications of the supported EFTs and bases, see the WCxf website at \url{https://wcxf.github.io/bases.html}).
The EFT and basis of a given \texttt{GlobalLikelihood} instance can be accessed via its \texttt{eft} and \texttt{basis} attributes.

\begin{tcolorbox}[breakable, size=fbox, boxrule=1pt, pad at break*=1mm,colback=cellbackground, colframe=cellborder]
\prompt{In}{incolor}{In}{\boxspacing}
\begin{Verbatim}[commandchars=\\\{\}]
\PY{n}{gl}\PY{o}{.}\PY{n}{eft}\PY{p}{,} \PY{n}{gl}\PY{o}{.}\PY{n}{basis}
\end{Verbatim}
\end{tcolorbox}

\begin{tcolorbox}[breakable, size=fbox, boxrule=.5pt, pad at break*=1mm, opacityfill=0]
\prompt{Out}{outcolor}{Out}{\boxspacing}
\begin{Verbatim}[commandchars=\\\{\}]
('SMEFT', 'Warsaw')
\end{Verbatim}
\end{tcolorbox}

\noindent
In order to create a likelihood function of Wilson coefficients in the WET, one can provide the \texttt{eft} and \texttt{basis} arguments on instantiation of a \texttt{GlobalLikelihood} instance.

\begin{tcolorbox}[breakable, size=fbox, boxrule=1pt, pad at break*=1mm,colback=cellbackground, colframe=cellborder]
\prompt{In}{incolor}{In}{\boxspacing}
\begin{Verbatim}[commandchars=\\\{\}]
\PY{n}{gl\PYZus{}wet} \PY{o}{=} \PY{n}{GlobalLikelihood}\PY{p}{(}\PY{n}{eft}\PY{o}{=}\PY{l+s+s1}{\PYZsq{}}\PY{l+s+s1}{WET}\PY{l+s+s1}{\PYZsq{}}\PY{p}{,} \PY{n}{basis}\PY{o}{=}\PY{l+s+s1}{\PYZsq{}}\PY{l+s+s1}{flavio}\PY{l+s+s1}{\PYZsq{}}\PY{p}{)}
\PY{n}{gl\PYZus{}wet}\PY{o}{.}\PY{n}{eft}\PY{p}{,} \PY{n}{gl\PYZus{}wet}\PY{o}{.}\PY{n}{basis}
\end{Verbatim}
\end{tcolorbox}

\begin{tcolorbox}[breakable, size=fbox, boxrule=.5pt, pad at break*=1mm, opacityfill=0]
\prompt{Out}{outcolor}{Out}{\boxspacing}
\begin{Verbatim}[commandchars=\\\{\}]
('WET', 'flavio')
\end{Verbatim}
\end{tcolorbox}

\subsubsection{Fixing a point in Wilson coefficient space: 3 equivalent ways}
\noindent
The point in the Wilson coefficient space at which the likelihood should be computed is defined using the \texttt{parameter\_point} method.
This method returns an instance of the \texttt{GlobalLikelihoodPoint} class that can be used to compute $\Delta \log{L}$.
The values of the Wilson coefficients can be provided in three equivalent ways:
\begin{itemize}

 \item A dictionary of Wilson coefficients as well as the scale in GeV at which they are defined can be passed directly as arguments.
\begin{tcolorbox}[breakable, size=fbox, boxrule=1pt, pad at break*=1mm,colback=cellbackground, colframe=cellborder]
\prompt{In}{incolor}{In}{\boxspacing}
\begin{Verbatim}[commandchars=\\\{\}]
\PY{n}{pp} \PY{o}{=} \PY{n}{gl}\PY{o}{.}\PY{n}{parameter\PYZus{}point}\PY{p}{(}\PY{p}{\PYZob{}}\PY{l+s+s1}{\PYZsq{}}\PY{l+s+s1}{lq3\PYZus{}2223}\PY{l+s+s1}{\PYZsq{}}\PY{p}{:} \PY{l+m+mf}{1e\PYZhy{}9}\PY{p}{\PYZcb{}}\PY{p}{,} \PY{n}{scale}\PY{o}{=}\PY{l+m+mi}{1000}\PY{p}{)}
\end{Verbatim}
\end{tcolorbox}

\item An instance of the \texttt{Wilson} class from the \texttt{wilson} package can be passed as a single argument.
    \begin{tcolorbox}[breakable, size=fbox, boxrule=1pt, pad at break*=1mm,colback=cellbackground, colframe=cellborder]
\prompt{In}{incolor}{In}{\boxspacing}
\begin{Verbatim}[commandchars=\\\{\}]
\PY{k+kn}{from} \PY{n+nn}{wilson} \PY{k+kn}{import} \PY{n}{Wilson}
\PY{n}{w} \PY{o}{=} \PY{n}{Wilson}\PY{p}{(}\PY{p}{\PYZob{}}\PY{l+s+s1}{\PYZsq{}}\PY{l+s+s1}{lq3\PYZus{}2223}\PY{l+s+s1}{\PYZsq{}}\PY{p}{:} \PY{l+m+mf}{1e\PYZhy{}9}\PY{p}{\PYZcb{}}\PY{p}{,} \PY{n}{scale}\PY{o}{=}\PY{l+m+mi}{1000}\PY{p}{,}
           \PY{n}{eft}\PY{o}{=}\PY{l+s+s1}{\PYZsq{}}\PY{l+s+s1}{SMEFT}\PY{l+s+s1}{\PYZsq{}}\PY{p}{,} \PY{n}{basis}\PY{o}{=}\PY{l+s+s1}{\PYZsq{}}\PY{l+s+s1}{Warsaw}\PY{l+s+s1}{\PYZsq{}}\PY{p}{)}
\PY{n}{pp} \PY{o}{=} \PY{n}{gl}\PY{o}{.}\PY{n}{parameter\PYZus{}point}\PY{p}{(}\PY{n}{w}\PY{p}{)}
\end{Verbatim}
\end{tcolorbox}

\item A WCxf file, e.g.\ a file in YAML format named \texttt{my\_wcxf.yaml} and containing
    \begin{Verbatim}[commandchars=\\\{\}]
eft: SMEFT
basis: Warsaw
scale: 1000
values:
  lq3\_2223:
    Re: 1e-9
    \end{Verbatim}

    can be read in by providing the path to the file as argument.

        \begin{tcolorbox}[breakable, size=fbox, boxrule=1pt, pad at break*=1mm,colback=cellbackground, colframe=cellborder]
\prompt{In}{incolor}{In}{\boxspacing}
\begin{Verbatim}[commandchars=\\\{\}]
\PY{n}{pp} \PY{o}{=} \PY{n}{gl}\PY{o}{.}\PY{n}{parameter\PYZus{}point}\PY{p}{(}\PY{l+s+s1}{\PYZsq{}}\PY{l+s+s1}{my\PYZus{}wcxf.yaml}\PY{l+s+s1}{\PYZsq{}}\PY{p}{)}
\end{Verbatim}
\end{tcolorbox}

\end{itemize}

\pagebreak

\subsubsection{Computing the likelihood}
\noindent
After the Wilson coefficients have been fixed and an instance of \texttt{GlobalLikelihoodPoint} has been created, it can be used to compute $\Delta \log{L}$.
In \texttt{smelli}, the global $\Delta \log{L}$ is given in terms of the sum of several individual $\Delta \log{L}$ that are constructed from subsets of observables.
To access all these individual $\Delta \log{L}$, the method \texttt{log\_likelihood\_dict} can be used.
It returns a dictionary containing the names of the individual likelihoods and the corresponding $\Delta \log{L}$ values.
Using the above defined parameter point, one gets

    \begin{tcolorbox}[breakable, size=fbox, boxrule=1pt, pad at break*=1mm,colback=cellbackground, colframe=cellborder]
\prompt{In}{incolor}{In}{\boxspacing}
\begin{Verbatim}[commandchars=\\\{\}]
\PY{n}{pp}\PY{o}{.}\PY{n}{log\PYZus{}likelihood\PYZus{}dict}\PY{p}{(}\PY{p}{)}
\end{Verbatim}
\end{tcolorbox}

            \begin{tcolorbox}[breakable, size=fbox, boxrule=.5pt, pad at break*=1mm, opacityfill=0]
\prompt{Out}{outcolor}{Out}{\boxspacing}
\begin{Verbatim}[commandchars=\\\{\}]
\{'fast\_likelihood\_quarks.yaml': 18.063309775625527,
 'fast\_likelihood\_leptons.yaml': -7.954151298861234e-05,
 'likelihood\_ewpt.yaml': 0.0019331634397694586,
 'likelihood\_eeww.yaml': -0.0001731988511934901,
 'likelihood\_lept.yaml': 3.7762380644679183e-07,
 'likelihood\_rd\_rds.yaml': 0.27864506193111893,
 'likelihood\_lfu\_fccc.yaml': 0.0005027179997831865,
 'likelihood\_lfu\_fcnc.yaml': 3.0607966063245655,
 'likelihood\_bcpv.yaml': 0.013775072147421241,
 'likelihood\_bqnunu.yaml': -0.119578242544371,
 'likelihood\_lfv.yaml': 0.0,
 'likelihood\_zlfv.yaml': 0.0,
 'likelihood\_higgs.yaml': 2.176258307784451e-05,
 'global': 21.299153554766516\}
\end{Verbatim}
\end{tcolorbox}

\noindent
%
%
While the global $\Delta \log{L}$ is provided by \texttt{log\_likelihood\_dict}, its value can also be directly returned using the \texttt{log\_likelihood\_global} method.

    \begin{tcolorbox}[breakable, size=fbox, boxrule=1pt, pad at break*=1mm,colback=cellbackground, colframe=cellborder]
\prompt{In}{incolor}{In}{\boxspacing}
\begin{Verbatim}[commandchars=\\\{\}]
\PY{n}{pp}\PY{o}{.}\PY{n}{log\PYZus{}likelihood\PYZus{}global}\PY{p}{(}\PY{p}{)}
\end{Verbatim}
\end{tcolorbox}

            \begin{tcolorbox}[breakable, size=fbox, boxrule=.5pt, pad at break*=1mm, opacityfill=0]
\prompt{Out}{outcolor}{Out}{\boxspacing}
\begin{Verbatim}[commandchars=\\\{\}]
21.299153554766516
\end{Verbatim}
\end{tcolorbox}

\noindent Apart from $\Delta \log{L}$, it is also possible to compute the total $\chi^2_{\rm NP}$, defined by
\begin{equation}
 \chi^2_{\rm NP} = -2\,\log{L_{\rm NP}}\,,
\end{equation}
where $L_{\rm NP}$ is normalized such that it is $1$ if the central values of the theory predictions are equal to the central values of the measurements for all observables.
A dictionary containing the individual values of the total $\chi^2_{\rm NP}$ is returned by the \texttt{chi2\_dict} method.

    \begin{tcolorbox}[breakable, size=fbox, boxrule=1pt, pad at break*=1mm,colback=cellbackground, colframe=cellborder]
\prompt{In}{incolor}{In}{\boxspacing}
\begin{Verbatim}[commandchars=\\\{\}]
\PY{n}{pp}\PY{o}{.}\PY{n}{chi2\PYZus{}dict}\PY{p}{(}\PY{p}{)}
\end{Verbatim}
\end{tcolorbox}

            \begin{tcolorbox}[breakable, size=fbox, boxrule=.5pt, pad at break*=1mm, opacityfill=0]
\prompt{Out}{outcolor}{Out}{\boxspacing}
\begin{Verbatim}[commandchars=\\\{\}]
\{'fast\_likelihood\_quarks.yaml': 160.14558316478963,
 'fast\_likelihood\_leptons.yaml': 23.57908813232271,
 'likelihood\_ewpt.yaml': 35.3618189920579,
 'likelihood\_eeww.yaml': 61.19130715429686,
 'likelihood\_lept.yaml': 1.4486600571844703,
 'likelihood\_rd\_rds.yaml': 34.10567278343568,
 'likelihood\_lfu\_fccc.yaml': 49.155325606131306,
 'likelihood\_lfu\_fcnc.yaml': 24.16370720780219,
 'likelihood\_bcpv.yaml': 5.140098429647292,
 'likelihood\_bqnunu.yaml': 21.417983245315177,
 'likelihood\_lfv.yaml': 8.998264557313096,
 'likelihood\_zlfv.yaml': -0.0,
 'likelihood\_higgs.yaml': 55.781752694208386,
 'global': 480.4892620245047\}
\end{Verbatim}
\end{tcolorbox}

\noindent
These values are particularly useful for computing p-values from the total $\chi^2_{\rm NP}$ and the number of observations.
The latter are returned by the \texttt{number\_observations\_dict} method of the \texttt{GlobalLikelihood} instance (which can be conveniently accessed using the \texttt{likelihood} attribute of the \texttt{GlobalLikelihoodPoint} instance).

    \begin{tcolorbox}[breakable, size=fbox, boxrule=1pt, pad at break*=1mm,colback=cellbackground, colframe=cellborder]
\prompt{In}{incolor}{In}{\boxspacing}
\begin{Verbatim}[commandchars=\\\{\}]
\PY{n}{pp}\PY{o}{.}\PY{n}{likelihood}\PY{o}{.}\PY{n}{number\PYZus{}observations\PYZus{}dict}\PY{p}{(}\PY{p}{)}
\end{Verbatim}
\end{tcolorbox}

            \begin{tcolorbox}[breakable, size=fbox, boxrule=.5pt, pad at break*=1mm, opacityfill=0]
\prompt{Out}{outcolor}{Out}{\boxspacing}
\begin{Verbatim}[commandchars=\\\{\}]
\{'fast\_likelihood\_quarks.yaml': 144,
 'fast\_likelihood\_leptons.yaml': 7,
 'likelihood\_ewpt.yaml': 30,
 'likelihood\_eeww.yaml': 48,
 'likelihood\_lept.yaml': 2,
 'likelihood\_rd\_rds.yaml': 11,
 'likelihood\_lfu\_fccc.yaml': 63,
 'likelihood\_lfu\_fcnc.yaml': 21,
 'likelihood\_bcpv.yaml': 6,
 'likelihood\_bqnunu.yaml': 22,
 'likelihood\_lfv.yaml': 41,
 'likelihood\_zlfv.yaml': 7,
 'likelihood\_higgs.yaml': 67,
 'global': 469\}
\end{Verbatim}
\end{tcolorbox}
\noindent
Note that here an ``observation'' is defined as an individual measurement of an observable. Thus, the number of observations is always greater than or equal to the number of observables.

\subsubsection{Table of observables}
\noindent
\texttt{smelli} provides information on individual observables. In particular, the theoretical and experimental central values and uncertainties as well as the pull compared to the SM or the experimental data can be obtained.
All this information is contained in an ``observable table'' that is returned in the form of a Pandas~\cite{reback2020pandas,mckinney-proc-scipy-2010} \texttt{DataFrame} object by the method \texttt{obstable}.

    \begin{tcolorbox}[breakable, size=fbox, boxrule=1pt, pad at break*=1mm,colback=cellbackground, colframe=cellborder]
\prompt{In}{incolor}{In}{\boxspacing}
\begin{Verbatim}[commandchars=\\\{\}]
\PY{n}{df} \PY{o}{=} \PY{n}{pp}\PY{o}{.}\PY{n}{obstable}\PY{p}{(}\PY{p}{)}
\end{Verbatim}
\end{tcolorbox}

\pagebreak

\noindent
In a Jupyter notebook, a Pandas \texttt{DataFrame} is shown as a table.

\begin{tcolorbox}[breakable, size=fbox, boxrule=1pt, pad at break*=1mm,colback=cellbackground, colframe=cellborder]
\prompt{In}{incolor}{In}{\boxspacing}
\begin{Verbatim}[commandchars=\\\{\}]
\PY{n}{df}
\end{Verbatim}
\end{tcolorbox}

\begin{tcolorbox}[breakable, size=fbox, boxrule=.5pt, pad at break*=1mm, opacityfill=0]
\prompt{Out}{outcolor}{Out}{\boxspacing}
\vskip5pt
\renewcommand{\arraystretch}{1.5}
\rowcolors{2}{white}{gray!10}
\resizebox{\textwidth}{!}{
\begin{tabular}{rrrrrrr}
  & \textbf{experiment} & \textbf{exp. unc.} & \textbf{theory} & \textbf{th. unc.} & \textbf{pull exp.} & \textbf{pull SM}\\
\hline
\textbf{a\_mu} & 0.00116592 & 6.31304e-10 & 0.00116592 & 4.25176e-10 & 3.49239 & -4.46085e-05 \\
\textbf{Rtaul(B->D*lnu)} & 0.296146 & 0.015608 & 0.244875 & 0 & 3.30606 & -0.389707 \\
\textbf{(<dR/dtheta>(ee->WW), 198.38, 0.8, 1.0)} & 6.535 & 0.236 & 7.236 & 0 & 2.97036 & 0.0112166 \\
\textbf{BR(W->taunu)} & 0.1138 & 0.0021 & 0.108417 & 0 & 2.56345 & -0.00503662 \\
\textbf{epsp/eps} & 0.00166382 & 0.000227703 & -3.12549e-05 & 0.000637111 & 2.50537 & 0.0147821 \\
\textbf{...} & ... & ... & ... & ... & ... & ... \\
\textbf{BR(tau->phie)} & 0 & 1.88467e-08 & 0 & 0 & 0 & 0\\
\textbf{BR(tau->phimu)} & 0 & 5.10684e-08 & 0 & 0 & 0 & 0\\
\textbf{BR(Z->emu)} & 0 & 2.33094e-07 & 0 & 0 & 0 & 0\\
\textbf{BR(Z->etau)} & 0 & 2.59807e-06 & 0 & 0 & 0 & 0\\
\textbf{BR(Z->mutau)} & 0 & 2.69574e-06 & 0 & 0 & 0 & 0\\
\end{tabular}
}
\vskip10pt
\scriptsize{399 rows $\times$ 6 columns}
\end{tcolorbox}

\noindent
The Pandas \texttt{DataFrame} is a convenient object for tabulated data and provides many useful features. E.g.\ one can sort the rows by the values of a given column,
%

\begin{tcolorbox}[breakable, size=fbox, boxrule=1pt, pad at break*=1mm,colback=cellbackground, colframe=cellborder]
\prompt{In}{incolor}{In}{\boxspacing}
\begin{Verbatim}[commandchars=\\\{\}]
\PY{n}{df}\PY{o}{.}\PY{n}{sort\PYZus{}values}\PY{p}{(}\PY{l+s+s1}{\PYZsq{}}\PY{l+s+s1}{pull SM}\PY{l+s+s1}{\PYZsq{}}\PY{p}{,} \PY{n}{ascending}\PY{o}{=}\PY{k+kc}{True}\PY{p}{)}\PY{p}{[}\PY{p}{:}\PY{l+m+mi}{5}\PY{p}{]}
\end{Verbatim}
\end{tcolorbox}

\begin{tcolorbox}[breakable, size=fbox, boxrule=.5pt, pad at break*=1mm, opacityfill=0]
\prompt{Out}{outcolor}{Out}{\boxspacing}
\vskip5pt
\renewcommand{\arraystretch}{1.5}
\rowcolors{2}{white}{gray!10}
\resizebox{\textwidth}{!}{
\begin{tabular}{rrrrrrr}
  & \textbf{experiment} & \textbf{exp. unc.} & \textbf{theory} & \textbf{th. unc.} & \textbf{pull exp.} & \textbf{pull SM}\\
\hline
\textbf{(<dBR/dq2>(Bs->phimumu), 1.0, 6.0)} & 2.55342e-08 & 3.72621e-09 & 4.04247e-08 & 6.44267e-09 & 2.0007 & -3.24157\\
\textbf{(<Rmue>(B0->K*ll), 1.1, 6.0)} & 0.681356 & 0.123108 & 0.746295 & 0 & 0.623038 & -2.4685\\
\textbf{BR(Bs->mumu)} & 2.73001e-09 & 3.80964e-10 & 2.73442e-09 & 1.47033e-10 & 0.0108006 & -2.29374\\
\textbf{(<dBR/dq2>(Bs->phimumu), 15.0, 19.0)} & 4.05106e-08 & 5.09449e-09 & 4.08896e-08 & 4.5361e-09 & 0.0555647 & -2.21418\\
\textbf{(<dBR/dq2>(B0->K*mumu), 15.0, 19.0)} & 4.35409e-08 & 3.61869e-09 & 4.35383e-08 & 6.16124e-09 & 0.000370693 & -2.20919\\
\end{tabular}
}
\end{tcolorbox}
\noindent
or select a specific row by its name.

    \begin{tcolorbox}[breakable, size=fbox, boxrule=1pt, pad at break*=1mm,colback=cellbackground, colframe=cellborder]
\prompt{In}{incolor}{In}{\boxspacing}
\begin{Verbatim}[commandchars=\\\{\}]
\PY{n}{df}\PY{o}{.}\PY{n}{loc}\PY{p}{[}\PY{p}{[}\PY{l+s+s1}{\PYZsq{}}\PY{l+s+s1}{Rtaul(B\PYZhy{}\PYZgt{}D*lnu)}\PY{l+s+s1}{\PYZsq{}}\PY{p}{]}\PY{p}{]}
\end{Verbatim}
\end{tcolorbox}

\begin{tcolorbox}[breakable, size=fbox, boxrule=.5pt, pad at break*=1mm, opacityfill=0]
\prompt{Out}{outcolor}{Out}{\boxspacing}
\vskip5pt
\renewcommand{\arraystretch}{1.5}
\rowcolors{2}{white}{gray!10}
\scriptsize{
\begin{tabular}{rrrrrrr}
  & \textbf{experiment} & \textbf{exp. unc.} & \textbf{theory} & \textbf{th. unc.} & \textbf{pull exp.} & \textbf{pull SM}\\
\hline
\textbf{Rtaul(B->D*lnu)} & 0.296146 & 0.015608 & 0.244875 & 0 & 3.30606 & -0.389707\\
\end{tabular}
}
\end{tcolorbox}

\pagebreak

\subsubsection{Plots}
\noindent
Given a likelihood function, one common task is to plot this function in a 2D plane.
In order to simplify this, \texttt{smelli} provides a method to compute the plot data for all individual likelihoods.
For demonstration, it is convenient to define a \texttt{GlobalLikelihood} instance for which the likelihood can be computed much faster than in the default case.
This can be achieved by considering only a subset of observables, e.g.\ only EWPO and the Higgs signal strengths.
    \begin{tcolorbox}[breakable, size=fbox, boxrule=1pt, pad at break*=1mm,colback=cellbackground, colframe=cellborder]
\prompt{In}{incolor}{In}{\boxspacing}
\begin{Verbatim}[commandchars=\\\{\}]
\PY{n}{gl\PYZus{}ewpt\PYZus{}higgs} \PY{o}{=} \PY{n}{GlobalLikelihood}\PY{p}{(}\PY{n}{include\PYZus{}likelihoods}\PY{o}{=}\PY{p}{[}
    \PY{l+s+s1}{\PYZsq{}}\PY{l+s+s1}{likelihood\PYZus{}ewpt.yaml}\PY{l+s+s1}{\PYZsq{}}\PY{p}{,}
    \PY{l+s+s1}{\PYZsq{}}\PY{l+s+s1}{likelihood\PYZus{}higgs.yaml}\PY{l+s+s1}{\PYZsq{}}\PY{p}{,}
\PY{p}{]}\PY{p}{)}
\end{Verbatim}
\end{tcolorbox}

\noindent
The next step is to define a function of the two plot parameters that returns a dictionary of Wilson coefficients.
This function defines what is actually plotted.
It can be a trivial function that takes two Wilson coefficients as arguments and just returns them, but it can also be a complicated function of two NP model parameters that returns a large set of Wilson coefficients depending on these two parameters.
As an example, we will reproduce figure~2 of~\cite{Falkowski:2019hvp} and plot the likelihood in the space of the $S$ and $T$ parameters.
They are proportional to the SMEFT Wilson coefficients $C_{\phi WB}$ and $C_{\phi D}$, and their relations are given by
\begin{equation}
 C_{\phi WB} = \frac{g_L\,g_Y}{16\,\pi\,v^2}\,S\,,
 \qquad
 C_{\phi D} = -\frac{g_L^2\,g_Y^2}{2\,\pi\,(g_L^2+g_Y^2)\,v^2}\,T .
\end{equation}
Consequently, plugging in the SM parameters, the function that takes $S$ and $T$ as arguments and returns a dictionary of Wilson coefficients can be defined as follows.

    \begin{tcolorbox}[breakable, size=fbox, boxrule=1pt, pad at break*=1mm,colback=cellbackground, colframe=cellborder]
\prompt{In}{incolor}{In}{\boxspacing}
\begin{Verbatim}[commandchars=\\\{\}]
\PY{k}{def} \PY{n+nf}{wc\PYZus{}fct}\PY{p}{(}\PY{n}{S}\PY{p}{,} \PY{n}{T}\PY{p}{)}\PY{p}{:}
    \PY{k}{return} \PY{p}{\PYZob{}}
        \PY{l+s+s1}{\PYZsq{}}\PY{l+s+s1}{phiWB}\PY{l+s+s1}{\PYZsq{}}\PY{p}{:} \PY{n}{S} \PY{o}{*} \PY{l+m+mf}{7.643950529889027e\PYZhy{}08}\PY{p}{,}
        \PY{l+s+s1}{\PYZsq{}}\PY{l+s+s1}{phiD}\PY{l+s+s1}{\PYZsq{}}\PY{p}{:} \PY{o}{\PYZhy{}}\PY{n}{T} \PY{o}{*} \PY{l+m+mf}{2.5793722852276787e\PYZhy{}07}\PY{p}{,}
    \PY{p}{\PYZcb{}}
\end{Verbatim}
\end{tcolorbox}

\noindent
This function can now be used as the first argument of the \texttt{plot\_data\_2d} method of the \texttt{GlobalLikelihood} instance.
The second argument is the scale at which the Wilson coefficients are defined, followed by the minimum and maximum values for the x- and y-axis.
In the function call below, also two optional arguments are given:
the number of \texttt{steps} in each direction (\texttt{steps} $=10$ results in plot data computed on a $10 \times 10$ grid), and the number of CPU threads to be used for the computation.

    \begin{tcolorbox}[breakable, size=fbox, boxrule=1pt, pad at break*=1mm,colback=cellbackground, colframe=cellborder]
\prompt{In}{incolor}{In}{\boxspacing}
\begin{Verbatim}[commandchars=\\\{\}]
\PY{n}{plot\PYZus{}data} \PY{o}{=} \PY{n}{gl\PYZus{}ewpt\PYZus{}higgs}\PY{o}{.}\PY{n}{plot\PYZus{}data\PYZus{}2d}\PY{p}{(}
    \PY{n}{wc\PYZus{}fct}\PY{p}{,}
    \PY{l+m+mf}{91.1876}\PY{p}{,}
    \PY{o}{\PYZhy{}}\PY{l+m+mf}{0.2}\PY{p}{,} \PY{l+m+mf}{0.2}\PY{p}{,} \PY{o}{\PYZhy{}}\PY{l+m+mf}{0.1}\PY{p}{,} \PY{l+m+mf}{0.3}\PY{p}{,}
    \PY{n}{steps}\PY{o}{=}\PY{l+m+mi}{10}\PY{p}{,}
    \PY{n}{threads}\PY{o}{=}\PY{l+m+mi}{8}\PY{p}{,}
\PY{p}{)}
\end{Verbatim}
\end{tcolorbox}

\noindent
The \texttt{plot\_data\_2d} method returns a dictionary with the names of the individual likelihoods as keys and values that are again dictionaries.
The keys in these latter dictionaries are \texttt{x}, \texttt{y}, and \texttt{z} and the values are arrays.
%
Here, \texttt{x} and \texttt{y} correspond to the coordinates in the 2D plane and \texttt{z} to the values of $\Delta \chi^2=-2\,\Delta \log{L}$ at these coordinates.
The dictionaries with keys \texttt{x}, \texttt{y}, and \texttt{z} are constructed in such a way that they can be directly fed  to the
\texttt{contour} plotting function of the \texttt{flavio} package.
%
The relevant submodules  for plotting have to be imported from \texttt{flavio} and \texttt{matplotlib}~\cite{Hunter:2007} (on which the \texttt{flavio} plotting functions are based on).

    \begin{tcolorbox}[breakable, size=fbox, boxrule=1pt, pad at break*=1mm,colback=cellbackground, colframe=cellborder]
\prompt{In}{incolor}{In}{\boxspacing}
\begin{Verbatim}[commandchars=\\\{\}]
\PY{k+kn}{import} \PY{n+nn}{flavio}\PY{n+nn}{.}\PY{n+nn}{plots} \PY{k}{as} \PY{n+nn}{fpl}
\PY{k+kn}{import} \PY{n+nn}{matplotlib}\PY{n+nn}{.}\PY{n+nn}{pyplot} \PY{k}{as} \PY{n+nn}{plt}
\end{Verbatim}
\end{tcolorbox}

\noindent
In order to plot $\Delta \chi^2$ contours corresponding to a given pull in units of $\sigma$, the contour levels can be defined using the \texttt{flavio} function \texttt{delta\_chi2}, which takes the number of $\sigma$ and the number of degrees of freedom as arguments.

    \begin{tcolorbox}[breakable, size=fbox, boxrule=1pt, pad at break*=1mm,colback=cellbackground, colframe=cellborder]
\prompt{In}{incolor}{In}{\boxspacing}
\begin{Verbatim}[commandchars=\\\{\}]
\PY{n}{levels\PYZus{}1sig} \PY{o}{=} \PY{p}{[}\PY{n}{fpl}\PY{o}{.}\PY{n}{delta\PYZus{}chi2}\PY{p}{(}\PY{l+m+mi}{1}\PY{p}{,} \PY{n}{dof}\PY{o}{=}\PY{l+m+mi}{2}\PY{p}{)}\PY{p}{]}
\PY{n}{levels\PYZus{}123sig} \PY{o}{=} \PY{p}{[}\PY{n}{fpl}\PY{o}{.}\PY{n}{delta\PYZus{}chi2}\PY{p}{(}\PY{n}{n\PYZus{}sigma}\PY{p}{,} \PY{n}{dof}\PY{o}{=}\PY{l+m+mi}{2}\PY{p}{)} \PY{k}{for} \PY{n}{n\PYZus{}sigma} \PY{o+ow}{in} \PY{p}{(}\PY{l+m+mi}{1}\PY{p}{,}\PY{l+m+mi}{2}\PY{p}{,}\PY{l+m+mi}{3}\PY{p}{)}\PY{p}{]}
\end{Verbatim}
\end{tcolorbox}

\noindent
The data can now be plotted. The function \texttt{fpl.contour} is called three times, once for each of the three different likelihoods: Higgs physics, EWPO, and their combination.
Furthermore, horizontal and vertical axes as well as labels are added.
A value larger than one for the argument \texttt{interpolation\_factor} of \texttt{fpl.contour} makes the contours appear smooth. However, if the plot data has been computed on a small grid, \texttt{interpolation\_factor} can obscure the fact that the data might be insufficient for a reasonable plot.
In fact, for more reasonable plots, the number of \texttt{steps} should be increased to at least $20$ (but this of course also increases the computing time).
From the data computed above, the plot is then generated by the following code.

    \begin{tcolorbox}[breakable, size=fbox, boxrule=1pt, pad at break*=1mm,colback=cellbackground, colframe=cellborder]
\prompt{In}{incolor}{In}{\boxspacing}
\begin{Verbatim}[commandchars=\\\{\}]
\PY{n}{plt}\PY{o}{.}\PY{n}{figure}\PY{p}{(}\PY{n}{figsize}\PY{o}{=}\PY{p}{(}\PY{l+m+mi}{5}\PY{p}{,}\PY{l+m+mi}{5}\PY{p}{)}\PY{p}{)}
\PY{n}{fpl}\PY{o}{.}\PY{n}{contour}\PY{p}{(}\PY{o}{*}\PY{o}{*}\PY{n}{plot\PYZus{}data}\PY{p}{[}\PY{l+s+s1}{\PYZsq{}}\PY{l+s+s1}{likelihood\PYZus{}higgs.yaml}\PY{l+s+s1}{\PYZsq{}}\PY{p}{]}\PY{p}{,} \PY{n}{levels}\PY{o}{=}\PY{n}{levels\PYZus{}1sig}\PY{p}{,}
            \PY{n}{label}\PY{o}{=}\PY{l+s+sa}{r}\PY{l+s+s2}{\PYZdq{}}\PY{l+s+s2}{Higgs (\PYZdl{}1}\PY{l+s+s2}{\PYZbs{}}\PY{l+s+s2}{sigma\PYZdl{})}\PY{l+s+s2}{\PYZdq{}}\PY{p}{,} \PY{n}{interpolation\PYZus{}factor}\PY{o}{=}\PY{l+m+mi}{9}\PY{p}{,}
            \PY{n}{color}\PY{o}{=}\PY{l+s+s1}{\PYZsq{}}\PY{l+s+s1}{C0}\PY{l+s+s1}{\PYZsq{}}\PY{p}{)}
\PY{n}{fpl}\PY{o}{.}\PY{n}{contour}\PY{p}{(}\PY{o}{*}\PY{o}{*}\PY{n}{plot\PYZus{}data}\PY{p}{[}\PY{l+s+s1}{\PYZsq{}}\PY{l+s+s1}{likelihood\PYZus{}ewpt.yaml}\PY{l+s+s1}{\PYZsq{}}\PY{p}{]}\PY{p}{,} \PY{n}{levels}\PY{o}{=}\PY{n}{levels\PYZus{}1sig}\PY{p}{,}
            \PY{n}{label}\PY{o}{=}\PY{l+s+sa}{r}\PY{l+s+s2}{\PYZdq{}}\PY{l+s+s2}{EWPO (\PYZdl{}1}\PY{l+s+s2}{\PYZbs{}}\PY{l+s+s2}{sigma\PYZdl{})}\PY{l+s+s2}{\PYZdq{}}\PY{p}{,} \PY{n}{interpolation\PYZus{}factor}\PY{o}{=}\PY{l+m+mi}{9}\PY{p}{,}
            \PY{n}{color}\PY{o}{=}\PY{l+s+s1}{\PYZsq{}}\PY{l+s+s1}{C1}\PY{l+s+s1}{\PYZsq{}}\PY{p}{)}
\PY{n}{fpl}\PY{o}{.}\PY{n}{contour}\PY{p}{(}\PY{o}{*}\PY{o}{*}\PY{n}{plot\PYZus{}data}\PY{p}{[}\PY{l+s+s1}{\PYZsq{}}\PY{l+s+s1}{global}\PY{l+s+s1}{\PYZsq{}}\PY{p}{]}\PY{p}{,} \PY{n}{levels}\PY{o}{=}\PY{n}{levels\PYZus{}123sig}\PY{p}{,}
            \PY{n}{label}\PY{o}{=}\PY{l+s+sa}{r}\PY{l+s+s2}{\PYZdq{}}\PY{l+s+s2}{global}\PY{l+s+s2}{\PYZdq{}}\PY{p}{,} \PY{n}{interpolation\PYZus{}factor}\PY{o}{=}\PY{l+m+mi}{9}\PY{p}{,}
            \PY{n}{color}\PY{o}{=}\PY{l+s+s1}{\PYZsq{}}\PY{l+s+s1}{C3}\PY{l+s+s1}{\PYZsq{}}\PY{p}{)}
\PY{n}{plt}\PY{o}{.}\PY{n}{axhline}\PY{p}{(}\PY{n}{c}\PY{o}{=}\PY{l+s+s1}{\PYZsq{}}\PY{l+s+s1}{0.6}\PY{l+s+s1}{\PYZsq{}}\PY{p}{,} \PY{n}{linewidth}\PY{o}{=}\PY{l+m+mi}{1}\PY{p}{)}
\PY{n}{plt}\PY{o}{.}\PY{n}{axvline}\PY{p}{(}\PY{n}{c}\PY{o}{=}\PY{l+s+s1}{\PYZsq{}}\PY{l+s+s1}{0.6}\PY{l+s+s1}{\PYZsq{}}\PY{p}{,} \PY{n}{linewidth}\PY{o}{=}\PY{l+m+mi}{1}\PY{p}{)}
\PY{n}{plt}\PY{o}{.}\PY{n}{xlabel}\PY{p}{(}\PY{l+s+sa}{r}\PY{l+s+s1}{\PYZsq{}}\PY{l+s+s1}{\PYZdl{}S\PYZdl{}}\PY{l+s+s1}{\PYZsq{}}\PY{p}{)}
\PY{n}{plt}\PY{o}{.}\PY{n}{ylabel}\PY{p}{(}\PY{l+s+sa}{r}\PY{l+s+s1}{\PYZsq{}}\PY{l+s+s1}{\PYZdl{}T\PYZdl{}}\PY{l+s+s1}{\PYZsq{}}\PY{p}{)}
\PY{n}{plt}\PY{o}{.}\PY{n}{legend}\PY{p}{(}\PY{p}{)}
\PY{n}{plt}\PY{o}{.}\PY{n}{show}\PY{p}{(}\PY{p}{)}
\end{Verbatim}
\end{tcolorbox}

\begin{center}
\includegraphics{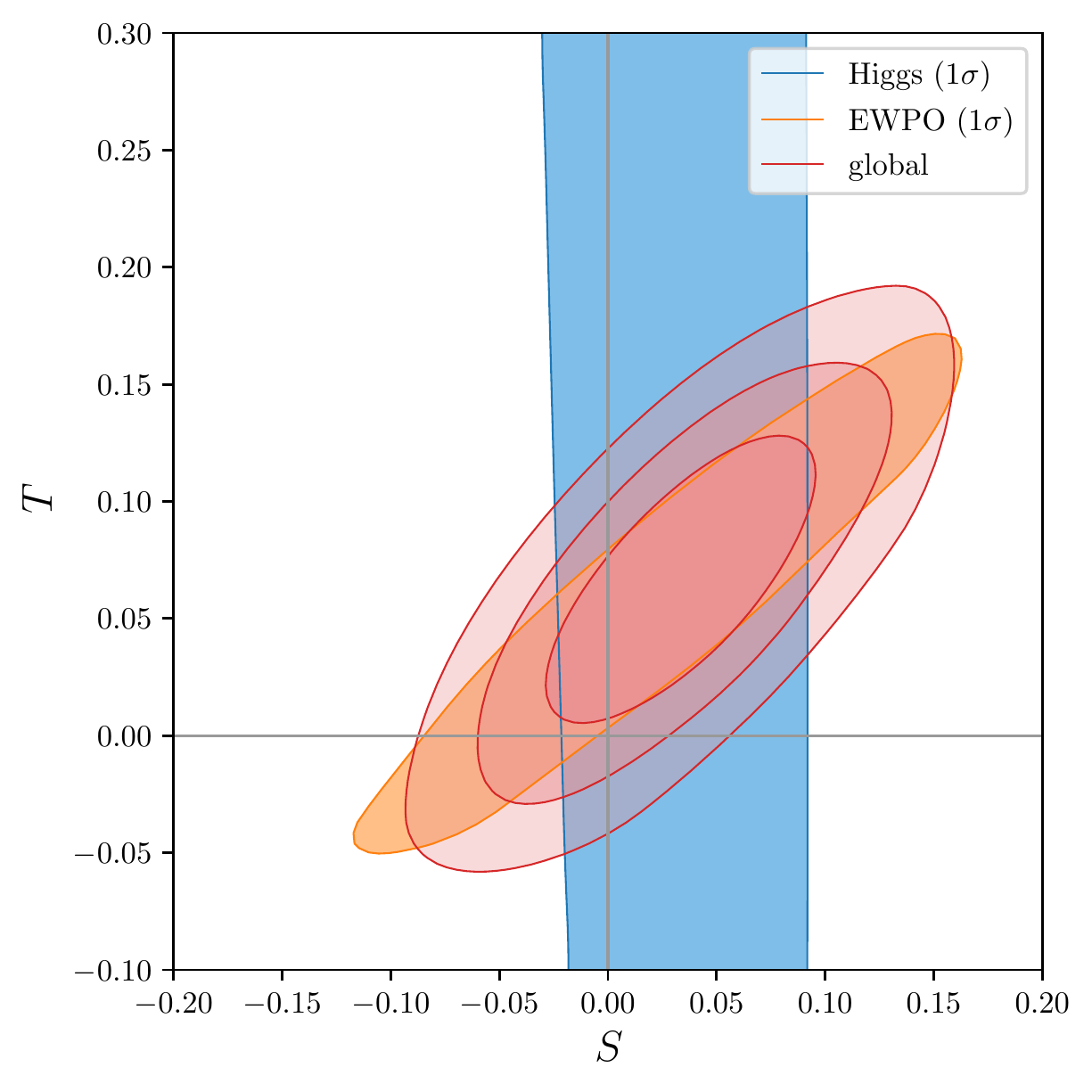}
\end{center}

\section{Conclusions}
\begin{itemize}
 \item Models that explain experimental deviations from the SM in certain observables generically predict also effects in other observables.
 This is e.g.\ the case for most models that explain the $B$~anomalies.
 Consequently, to test such models, one has to consider a \emph{global} likelihood constructed from as many observables as possible.
 \item This article shows how to use the python package \texttt{smelli}, which implements a global \underline{SME}FT \underline{l}ike\underline{li}hood function.
 It can be used to either test models, or to interpret data model-independently in the WET and the SMEFT.
 To date, 399 flavor and other precision observables are included in the likelihood.
 \item The full \emph{global} likelihood is work in progress. Since \texttt{smelli} is completely open source, you are welcome to join us on \url{https://github.com/smelli/smelli} and to participate in the effort to make \texttt{smelli} truly global.
\end{itemize}

\section*{Acknowledgments}
I thank Jason Aebischer, Jacky Kumar, and David M. Straub for the collaboration \texttt{smelli} is based on and Matthew Kirk for contributing to \texttt{smelli} on GitHub.

\bibliographystyle{JHEP}
\bibliography{bibliography}

\end{document}